\documentclass[12pt,a4paper]{article}
\usepackage[latin1]{inputenc}
\usepackage[a4paper]{geometry}
\usepackage{graphicx}
\usepackage{caption}
\usepackage{hyperref}
\usepackage{setspace}
\usepackage{amsmath}
\usepackage{amssymb}
\usepackage{amsthm}
\usepackage{float}
\usepackage{times}
\usepackage{nicefrac}
\usepackage{color}
\usepackage{xcolor}

\hypersetup{
	colorlinks,
	linkcolor={red!50!black},
	citecolor={blue!50!black},
	urlcolor={blue!80!black}
}

\begin{document}

\title{An Inverse-Ramsey Tax Rule}

\author{Luca Micheletto\thanks{Department of Law, University of Milan, Italy. Email: luca.micheletto@unimi.it } \and Dylan Moore\thanks{Department of Economics, University of Hawaii, Honolulu, HI, USA. Email: dtmoore@hawaii.edu} \and Daniel Reck\thanks{Department of Economics, University of Maryland, College Park, MD, USA. Email: dreck@umd.edu} \and Joel Slemrod\thanks{
		Department of Economics, University of Michigan, Ann Arbor, MI, USA. Email:
		jslemrod@umich.edu} }

\maketitle
\begin{abstract}
\onehalfspacing

Traditional optimal commodity tax analysis, dating back to Ramsey (1927), prescribes that to maximize welfare one should impose higher taxes on goods with lower demand elasticities. Yet policy makers do not stress minimizing efficiency costs as a desideratum. In this note we revisit the commodity tax problem, and show that the attractiveness of the Ramsey inverse-elasticity prescription can itself be inverted if the tax system is chosen -- or at least strongly influenced -- by taxpayers who are overly confident of their ability, relative to others, to substitute away from taxed goods.

\smallskip

\textbf{Keywords:} Commodity taxation, overconfidence, behavioral anomalies

\smallskip

\textbf{JEL classification:} H20, D90 
\end{abstract}

\onehalfspacing

\newpage{}

\section{Introduction and motivation}

Imagine that a taxpayer can choose between two new taxes, each levied on a previously untaxed good with the same expenditure. Which should they prefer?  At first blush, they should prefer a tax on the good for which they feel there are suitable alternatives available, because a tax on that good will reduce their utility less.

A thoughtful, benevolent tax policy maker, perhaps familiar with the theory of optimal commodity taxation, would overrule the taxpayer's instinct, realizing that the taxpayer-preferred tax will yield less revenue precisely because of the relatively elastic behavioral response. Upon careful reflection of which tax best serves the taxpayer's interest while raising a given level of revenue, the policy maker would reverse the taxpayer's initial preference and tax the good with a lower demand elasticity, that is the good for which the taxpayer has less good alternatives.

This benevolent policy maker would have followed one of the most venerable optimal tax prescriptions, known as the Ramsey rule, first formulated by Frank Ramsey in an article published in 1927.  It states that, in an optimal commodity tax system with identical consumers, every taxed good would have the same proportional reduction in compensated demand. This implies that goods with more elastic demand should be taxed at a lower rate. Moreover, if cross-elasticities of demand are zero, it leads to an inverse elasticity rule, under which the optimal tax on a good is inversely proportional to its own demand elasticity. Although our opening example limited the policy choice to taxing only one good, more generally efficiency considerations -- minimizing excess burden -- imply that taxing inelastic bases is an attractive rule of thumb. In a world where there is a revenue requirement, the behavioral response that the taxpayer sees as a way to reduce her own burden is the very source of inefficiency.

In spite of the respect that taxing inelastic bases attracts among academic economists, many observers have noted that efficiency in this sense figures hardly at all in voters' views about taxation and in the politics, as opposed to the economics, of taxation.  Policy debate might consider how behavioral response will affect the likely amount of revenue raised, but will generally not link that to an efficiency argument. This is the conclusion of Stantcheva (2021) who, based on large-scale social economics surveys and experiments, finds that efficiency arguments play no role in Americans' views about income and estate taxation. An extreme version of this view was taken by Tanzi (2008), who argued that ``optimal taxation...has been largely ignored'' and that ``its recommendations often conflict with what governments want to do or what taxpayers expect them to do.''  Boadway (2012) offers an impassioned rebuttal. 

In this note, we propose a behavioral model that explains why tax policy often disregards the Ramsey-style reasoning based on how the burden of taxes relative to revenue raised is perceived by taxpayers. The key feature of the model is that taxpayers are overconfident about their ability, relative to their peers, to escape taxation by substituting away from taxed goods: they believe that they have a larger elasticity than others. We show that, in this scenario, the optimal tax regime can feature what we call an ``inverse-Ramsey'' tax rule, where tax rates are higher for goods with relatively elastic bases. Intuitively, taxpayers may prefer the tax on the elastic base because they over-estimate their relative ability to avoid the tax and thereby reduce their burden. In certain settings, such overconfidence can also lead taxpayers to prefer subsidies on some goods, and even to prefer a distorting commodity tax system to a non-distorting lump-sum tax. 

This approach has implications beyond commodity taxation for the political equilibrium of tax policy. In particular, it can explain the widespread public opposition to increases in taxes whose base is largely or completely unaffected by current or future decisions.  In this category is the estate tax, whose (limited) popularity is addressed in Slemrod (2006), and the base of which is past wealth accumulation decisions.  A similar argument applies to wealth taxes. Somewhat more subtly, the argument applies to the transition from an income tax to a consumption tax, under which all income previously saved, and presumably taxed, under the income tax would be subject to tax again when consumed, absent complex transition rules. More generally, as considered by Levmore (1993), retroactive taxes by definition are based on past, and therefore immutable, decisions; they are notably disliked. Most directly, a lump-sum tax depends on no decisions (except perhaps migration), and its popularity is so self-evidently meager (undoubtedly largely due to its distributional impotence) that to our knowledge it has never been measured in surveys. All of these taxes fare well in standard optimal tax models that seek to minimize the social cost of distorted choices, because much or all of the tax base is not affected by personal choices such as consumption or labor supply. They generally meet with widespread popular and political push-back. What they share is that they do not draw support from taxpayers who are overly confident of their ability to take advantage of the possibility of tax-reducing behavioral response.

\section{An inverse-Ramsey rule deriving from overconfidence\label{Section2}}

In this section, we present the model that can generate an inverse-Ramsey rule for optimal commodity taxation: a model where the social planner taxes more highly those goods they perceive to be more elastically demanded. Because we focus on the consumer side of the economy, we follow the bulk of the
literature in making the simplest possible assumptions about production.\footnote{See, for instance, Atkinson and Stiglitz (1972, 1976).} In
particular, we assume that $n$\ commodities are produced by a linear
technology using labor as the only input, and that units are
chosen to make all producer prices equal to 1. Consumer prices are denoted
by $q_{i}=1+t_{i}$, for $i=1,...,n$, with $t_{i}$ being the tax rate on
good $i$.

The total size of the population is normalized to 1, and it consists of a
large number of identical agents whose preferences are represented by a
quasi-concave utility function $u\left( x_{1},...,x_{n},L\right) $, where $L$
denotes labor supply and $x_{i}$, for $i=1,...,n$, denotes the consumption of
commodity $i$. The workers' wage rate is denoted by $w$ and labor income is
assumed to be the only source of purchasing power for the individuals.
Finally, with demand functions being homogeneous of degree zero in consumer
prices, we assume without loss of generality that labor income goes untaxed. Denote by $\mathbf{q}$ the vector $\left( q_{1},...,q_{n}\right) $, and by $%
\mathbf{x}$ the vector $\left( x_{1},...,x_{n}\right) $. For given tax
rates, each agent makes choices that maximize $u\left( \mathbf{x},L\right) $
subject to the private budget constraint $\mathbf{q}\cdot \mathbf{x}=wL$.
This generates the demand vector $\mathbf{x}^{\ast }\left( \mathbf{q}%
,w\right) $ with associated (indirect) utility $V\left( \mathbf{q},w\right)
\equiv u\left( \mathbf{x}^{\ast }\left( \mathbf{q},w\right) ,\frac{\mathbf{qx%
}^{\ast }\left( \mathbf{q},w\right) }{w}\right) $.

Despite the fact that agents are fully rational when acting as
decision-makers in the various markets (i.e., as suppliers in the labor
market, and as consumers in the commodity markets), we assume that, individually, each of
them may misperceive the behavior of other agents. Thus, assuming that tax revenues are needed to finance an exogenous revenue
requirement denoted by $R$, the tax structure that would be favored by a
representative agent or voter\footnote{Note that, assuming a population of identical voters satisfying this description,
our planner's problem follows immediately as a characterization of
the policy equilibrium in both a median voter model and a probabilistic
voting model.} can be described as the outcome of the following
\begin{equation*}
\underset{t_{1},...,t_{n}}{\max }V\left( q_{1},...,q_{n},w\right) 
\end{equation*}
\newline
subject to
\begin{equation}
\sum_{k}t_{k}\overline{x}_{k}\geq R,  \label{1}
\end{equation}
\newline
where the vector $\left( \overline{x}_{1},...,\overline{x}_{n}\right) $
describes a representative agent's beliefs about the amount of the
various goods that are consumed by the other members of society (for a given consumer-price vector $\mathbf{q}$). Because all agents are assumed to be identical, they are also
assumed to hold the same beliefs. These beliefs can, however, be wrong.\footnote{Note that, although our model can be regarded as \textquotedblleft
behavioral\textquotedblright in the sense that people may have systematic
misperceptions that affect what they believe to be the optimal tax
structure, it does not fit within the general framework for optimal taxation
with behavioral agents presented by Farhi and Gabaix (2020). The main reason
is that, in their framework, agents are \textit{not} fully rational when
acting as decision makers in the various markets (for instance, because they
misperceive some of the price signals). Put differently, within their
framework the demand vector that we have denoted by $\mathbf{x}^{\ast
}\left( \mathbf{q},w\right) $, which describes the market-behavior of
agents, would \textit{not} result from the maximization of the
\textquotedblleft true\textquotedblright\ utility $u\left( \mathbf{x}
,L\right) $ subject to the budget constraint $\mathbf{q}\cdot \mathbf{x}=wL$
. Another difference between our model and Farhi and Gabaix
(2020) is that their analysis is normative in nature, whereas ours falls
within the realm of political economy--what policy will emerge, as opposed to what policy should emerge.}
Formally, denoting by $x_{i}^{\ast }$ the amount actually consumed of good $i$ by a
representative agent, wrong beliefs imply that $x_{i}^{\ast }/\overline{x}
_{i}\neq 1$ for some $i\in \left\{ 1,...,n\right\} $.

We propose that wrong beliefs originate from a confidence
bias. Individuals are assumed to be overconfident in the sense that they
overestimate their own (relative) ability to escape the burden of taxation
by substituting away from relatively highly taxed goods. We also assume individuals correctly perceive $\overline{x}_{i}$ when good $i$ is not taxed.\footnote{This covers both the case where the tax on good $i$ is zero and cases where good $i$ is subsidized. For further discussion of this assumption and its implications, see footnote \ref{fn:supercases}.} With this form of overconfidence, for $t_i>0$, we have
$x_{i}^{\ast }/\overline{x}_{i}<1$.  Assuming that the higher the
own-price elasticity of a good, the larger is the extent of the individuals' confidence bias,
it follows that the ratio $x_{i}^{\ast }/\overline{x}_{i}$ is (weakly)
decreasing in the absolute value of the own-price elasticity of a good.

There is considerable evidence that overconfidence is a pervasive phenomenon. One popular introductory psychology textbook (Myers, 1998, p. 440) puts it as follows: "For nearly any subjective or socially desirable dimension...most people see themselves as better than average." A prominent survey of the intersection of psychology and economics (DellaVigna, 2009) addresses overconfidence as its first example of nonstandard beliefs, and argues that it helps explain behavior in a variety of settings. We note that the literature often distinguishes between what is called  overestimation, believing that one is better than reality justifies, and overplacement--believing, without support, that one is better than others. Here, we assume overplacement.

Some of the literature examines whether overplacement is more or less likely for easy versus difficult tasks, with most studies concluding that it is more likely for easy tasks (Kruger, 1999; Larrick, Burson, and Scott, 2007; Moore and Healey, 2008). Interpreting the own-price demand elasticity of a good as reflecting the ease of finding substitutes for its consumption, then the greater likelihood of overplacement in easier tasks aligns with our model assumption that the higher the own-price elasticity of a good, the larger the individual's bias.

The first-order condition for an interior solution to the perceived optimum for the tax rate on good $i$ in this model is
\begin{equation}
\underbrace{-\frac{\alpha}{\mu}\frac{x_{i}^{\ast}}{\overline{x}_{i}}}_{\text{private losses}}+\underbrace{1-\frac{t_{i}}{1+t_{i}}\overline{e}_{i}}_{\text{revenue effects}}=0,\label{first_order_condition}
\end{equation}
where $\overline{e}_{i}\equiv\vert\frac{\partial\overline{x}_{i}}{\partial q_{i}}\frac{q_{i}}{\overline{x}_{i}}\vert$ is perceived elasticity of demand for good $i$, and $\mu$ and $\alpha$ are the Lagrange multipliers for the consumer and perceived planner problems, respectively. 
This presentation of the first-order condition reflects the costs and benefits associated with increasing
the tax burden on good $i$ by \$1. We observe two key differences between equation \eqref{first_order_condition}
and the standard first-order condition in the Ramsey model. First, misperception
of the elasticity leads to erroneous beliefs about revenue effects,
as the expected behavioral response to taxation depends on the perceived
elasticity. Second, the social value of the private utility losses
associated with a \$1 increase in tax burden -- usually simply $\frac{\alpha}{\mu}$ -- is
modified by the multiplier $\frac{x_{i}^{\ast}}{\overline{x}_{i}}$. Intuitively, the term $\frac{x_{i}^{\ast}}{\overline{x}_{i}}<1$ is the representative agent's perception of their relative share of the burden of a tax on good $i$.

These two misperception effects work in concert to increase
the perceived attractiveness of a tax on good $i$.
Underestimating the elasticity of their peers makes the agents believe
both that the behavioral response to taxing good $i$ is smaller,
and that their own share of the burden of a tax on good $i$ is
smaller. The lower they perceive the responsiveness of others to be,
the more efficient they believe the tax is, and the more they believe
that the burden of the tax falls disproportionately on others.

Combining the first-order conditions for good $i$ and $j$ together,
we obtain a variant of the standard Ramsey condition:
\begin{equation}
\frac{\frac{t_{i}}{1+t_{i}}}{\frac{t_{j}}{1+t_{j}}}=\frac{1-\frac{\alpha}{\mu}\frac{x_{i}^{\ast}}{\overline{x}_{i}}}{1-\frac{\alpha}{\mu}\frac{x_{j}^{\ast}}{\overline{x}_{j}}}\frac{\overline{e}_{j}}{\overline{e}_{i}}.\label{2}
\end{equation} Even though the perceived elasticities do not necessarily coincide with true elasticities (given, respectively
for good $i$ and $j$, by $e_{i}\equiv\left\vert \frac{\partial x_{i}^{\ast}}{\partial q_{i}}\frac{q_{i}}{x_{i}^{\ast}}\right\vert $
and $e_{j}\equiv\left\vert \frac{\partial x_{j}^{\ast}}{\partial q_{j}}\frac{q_{j}}{x_{j}^{\ast}}\right\vert $), we assume that the ranking of perceived elasticities is consistent with that of true elasticities: if $e_{j}<\left(>\right)e_{i}$, then $\overline{e}_{j}<\left(>\right)\overline{e}_{i}$.  

It is easier to characterize when an inverse-Ramsey rule obtains if we impose a little more structure on the two related misperceptions in the model. For the remainder of this paper, we will focus on a specific case
where eq. (\ref{2}) can deliver an inverse-Ramsey outcome
(although there are others). In particular, we suppose that demand for good $j$
is correctly perceived by agents ($x_{j}^{\ast}=\overline{x}_{j}$
and $\overline{e}_{j}=e_{j}$), but people are overconfident about the response to taxation of good
$i$ ($\overline{e}_{i}<e_{i}$). We also assume that $e_{j}<e_{i}$ and that agents are
not so overconfident as to overturn the relative ranking of goods
in terms of perceived elasticities, i.e., $e_{j}<\overline{e}_{i}<e_{i}$.

In this setting, eq. (\ref{2}) simplifies to 
\begin{equation}
\frac{\frac{t_{i}}{1+t_{i}}}{\frac{t_{j}}{1+t_{j}}}=\frac{1-\frac{\alpha}{\mu}\frac{x_{i}^{\ast}}{\overline{x}_{i}}}{1-\frac{\alpha}{\mu}}\frac{e_{j}}{\overline{e}_{i}}.\label{eq:relTaxRates}
\end{equation}
If the only effect of overconfidence was to alter the relative magnitude
of the perceived elasticities of these two goods, this would
be insufficient to deliver an inverse Ramsey result, as we are still
assuming $i$ is perceived to be the more elastic good. However, the
fact that agents believe that the burden of a tax on good $i$ will be borne disproportionately
by others adds an additional force favoring taxation of good
$i$ over good $j$. It is this additional effect of misperception
that allows for the possibility that the planner will choose to place
a higher tax on the good that is (in both reality and perception)
relatively more elastic. That is, the planner may impose rates such
that the inverse of the Ramsey inverse-elasticity rule holds: $\frac{t_{i}}{1+t_{i}}>\frac{t_{j}}{1+t_{j}}$.

To see that this can indeed occur in some cases, note that equation
(\ref{eq:relTaxRates}) implies that an inverse-Ramsey result holds
if and only if
\begin{equation}
\frac{1-\frac{\alpha}{\mu}\frac{x_{i}^{\ast}}{\overline{x}_{i}}}{1-\frac{\alpha}{\mu}}e_{j}>\overline{e}_{i}.\label{condition}
\end{equation}
Condition \eqref{condition} is trivially satisfied when $\overline{e}_{i}=e_{j}$
(because $\frac{x_{i}^{\ast}}{\overline{x}_{i}}<1$). Continuity of
equation \eqref{condition} thus implies that there always exists
some sufficiently severe level of misperception which still satisfies
$\overline{e}_{i}>e_{j}$, but for which an inverse-Ramsey result
obtains.\footnote{Appendix \ref{existence} presents this result more formally, showing
that there always exists some perceived demand functions satisfying
$\overline{e}_{i}>e_{j}$, which deliver an inverse-Ramsey outcome.}

When $\overline{e}_{i}>e_{j}$, condition \eqref{condition} requires
that the ratio $x_{i}^{\ast}/\overline{x}_{i}$ is sufficiently small,
which in turn depends on how severely the elasticity $e_{i}$ is misperceived
(as a larger discrepancy between $e_{i}$ and $\overline{e}_{i}$
is likely to translate into a lower ratio $x_{i}^{\ast}/\overline{x}_{i}$).
Two further aspects of condition \eqref{condition} are worth noting.
First, it is not as simple as it might first appear, as $x_{i}^{\ast}/\overline{x}_{i}$
depends on the tax rates in place. Second, a tax rate structure that
meets the first-order conditions \eqref{first_order_condition} and
the perceived government budget constraint will not raise the required
revenue, because the chosen tax rates are
based on misperceived elasticities.

This second issue can be resolved by specifying a budget adjustment rule -- the process by which the decision-maker satisfies the government budget constraint despite their misperceptions.  In the literature on misperceptions (see, e.g., Chetty, Looney, and Kroft, 2009; Farhi and Gabaix, 2020), it is common to specify the decision-maker's choice as the option that satisfies the misperceived first-order condition \eqref{first_order_condition} and the \emph{actual} budget constraint. The intuition that justifies this rule is that our decision-maker may adjust to budget shortfalls as shocks to the revenue requirement (analogously to an income shock in consumer choice theory; see Reck (2016) for a more detailed discussion). If our decision-maker uses such a budget adjustment rule and condition \eqref{condition} is satisfied at the actual budget constraint, then the tax rates they select will be an inverse-Ramsey rule. Adopting this budget adjustment rule carries two caveats. First, we require that budget adjustment does not induce learning, e.g. that realizing there is a budget shortfall leads the decision-maker to correct her perceptions. Second, we assume it is feasible to satisfy the perceived first-order condition and the true budget constraint despite the existence of revenue-maximizing tax rates.\footnote{This assumption is unnecessary in prior work adopting this budget adjustment rule to study the typical consumer choice problem. This is because the consumer's budget constraint is linear in choice variables, while the government budget constraint is nonlinear in tax rates.}

\bigskip 

\section{An illustrative example}

We now consider a simpler illustrative example with two goods and a representative
consumer who maximizes a quasilinear utility function of the following form:
\[
u\left(x_{1},x_{2}\right)\equiv\frac{x_{1}^{1-\frac{1}{e_{1}}}}{1-\frac{1}{e_{1}}}+\frac{x_{2}^{1-\frac{1}{e_{2}}}}{1-\frac{1}{e_{2}}}-\left(1+ t_{1}\right)x_{1}-\left(1+ t_{2}\right)x_{2}
\]
where $ t_{i}\in\left(-1,\infty\right)$ is the tax rate on good
$i$ and $e_{i}>0$ is the elasticity of demand for good $i$. This
implies that the consumer's demand for good $i$ is $x_{i}^{*}=\left(1+ t_{i}\right)^{-e_{i}}$.

The decision-maker sets the two tax rates to maximize utility for
the representative agent subject to a revenue requirement, but misperceives
the budget constraint as
\[
 t_{1}\overline{x}_{1}+ t_{2}\overline{x}_{2}=R
\]
where $\overline{x}_{i}\equiv\left(1+ t_{i}\right)^{-\overline{e}_{i}}$
is the perceived demand for good $i$. Any difference between true
demand and perceived demand stems from incorrectly assuming the elasticity
of demand for good $i$ is $\overline{e}_{i}$ when in fact $\overline{e}_{i}\neq e_{i}$.

Now assume that $\overline{e}_{i}\equiv\theta_{i}e_{i}$, where $\theta_{i}$ captures the proportional misperception of others' elasticities (with a larger misperception being reflected in a lower $\theta _{i}$), with $\theta _{i}\in (0,1]$ for $ t _{i}>0$ and $\theta _{i}=1$ otherwise.\footnote{Notice that, even though the function $\theta _{i}\left( t_{i}\right) $ is
modelled as a step function (and therefore is potentially discontinuous at $
t_{i}=0$), the ensuing demand function $\overline{x}_{i}\left( t_{i}\right) $
is continuous (given that the value of $\overline{x}_{i}\left( t_{i}\right) $
does not depend on $\theta _{i}$ when $t_{i}=0$).}\textsuperscript{,}\footnote{\label{fn:supercases}Here, as in the previous section, we maintain the assumption that overconfidence
occurs only in the domain of goods that are \textit{taxed}. 
Relaxing this assumption may, under some circumstances, affect our results.
Suppose, for instance, that overconfidence is treated as symmetric, in the
sense that individuals are overly confident both of their ability to
substitute away from taxed goods and of their ability to substitute into
subsidized goods. That is, suppose that $\theta _{2}\in \left( 0,1\right) $
whether $ t _{2}$ is positive or negative. In this case the
ratio $\overline{x}_{2}/x_{2}$ is increasing in $t_{2}$ for $t_{2}>0$ but is
decreasing in $t_{2}$ for $t_{2}<0$. An inverse-Ramsey outcome may then be
dominated by a \textquotedblleft super-Ramsey\textquotedblright\ outcome
where the more elastic good (good 2 in our example) is subsidized and the
less elastic good is taxed. The intuition comes from the fact that each
individual believes that, at the same time, she would suffer a relatively
low burden (compared to others) when a tax is levied on the consumption of
good 2, and would enjoy a relatively large benefit when the consumption of
good 2 is subsidized. Thus, she may prefer to raise more than the
required revenue target by taxing good 1 in order to finance a subsidy on
good 2, rather than to raise more than the required revenue target by taxing
good 2 in order to finance a subsidy on good 1.}  Using
the true and perceived demand functions, we can write the decision-maker's
perceived problem as
\[
\max_{ t_{1}, t_{2}}\left\{ -\frac{\left(1+ t_{1}\right)^{1-e_{1}}}{1-e_{1}}-\frac{\left(1+ t_{2}\right)^{1-e_{2}}}{1-e_{2}}: t_{1}\left(1+ t_{1}\right)^{-\theta_{1}e_{1}}+ t_{2}\left(1+ t_{2}\right)^{-\theta_{2}e_{2}}=R\right\} .
\]
To describe the tax rates the decision-maker will choose, it is helpful to
consider the welfare effect of reforms that hold perceived revenue
constant. Let
\[
R^{s}\left( t_{1}, t_{2};\theta_{1},\theta_{2}\right)\equiv t_{1}\left(1+ t_{1}\right)^{-\theta_{1}e_{1}}+ t_{2}\left(1+ t_{2}\right)^{-\theta_{2}e_{2}}
\]
be perceived revenue. For a given increase in $ t_{1}$, the planner
believes holding revenue constant requires increasing $ t_{2}$
by 
\begin{align}
\left.\frac{\text{d} t_{2}}{\text{d} t_{1}}\right|_{R^{s}} & =-\frac{\nicefrac{\partial R^{s}}{\partial t_{1}}}{\nicefrac{\partial R^{s}}{\partial t_{2}}}\\
 & =-\frac{\left(1+ t_{1}\right)^{-\theta_{1}e_{1}}\left(1-\theta_{1}e_{1}\frac{ t_{1}}{1+ t_{1}}\right)}{\left(1+ t_{2}\right)^{-\theta_{2}e_{2}}\left(1-\theta_{2}e_{2}\frac{ t_{2}}{1+ t_{2}}\right)}.
\end{align}
The welfare effect of this reform is
\begin{align*}
\left.\frac{\text{d}\mathcal{W}}{\text{d} t_{1}}\right|_{R^{s}} & =\frac{\partial\mathcal{W}}{\partial t_{1}}+\frac{\partial\mathcal{W}}{\partial t_{2}}\cdot\left.\frac{\text{d} t_{2}}{\text{d} t_{1}}\right|_{R^{s}}\\
 & =\left(1+ t_{1}\right)^{-e_{1}}\left[\frac{\left(1+ t_{1}\right)^{\left(1-\theta_{1}\right)e_{1}}\left(1-\theta_{1}e_{1}\frac{ t_{1}}{1+ t_{1}}\right)}{\left(1+ t_{2}\right)^{\left(1-\theta_{2}\right)e_{2}}\left(1-\theta_{2}e_{2}\frac{ t_{2}}{1+ t_{2}}\right)}-1\right]
\end{align*}
Thus, the planner's perceived FOC
\[
\left.\frac{\text{d}\mathcal{W}}{\text{d} t_{1}}\right|_{R^{s}}=0
\]
is equivalent to
\begin{equation}
\left(1+ t_{1}\right)^{\left(1-\theta_{1}\right)e_{1}}\left(1-\theta_{1}e_{1}\frac{ t_{1}}{1+ t_{1}}\right)=\left(1+ t_{2}\right)^{\left(1-\theta_{2}\right)e_{2}}\left(1-\theta_{2}e_{2}\frac{ t_{2}}{1+ t_{2}}\right).\label{eq:plannersFOC}
\end{equation}
This equation, together with the perceived budget constraint
\begin{equation}
 t_{1}\left(1+ t_{1}\right)^{-\theta_{1}e_{1}}+ t_{2}\left(1+ t_{2}\right)^{-\theta_{2}e_{2}}=R,\label{eq:percBC}
\end{equation}
characterize the solution to the decision-maker's perceived problem.

\begin{figure}
\begin{centering}
\includegraphics[width=0.8\paperwidth]{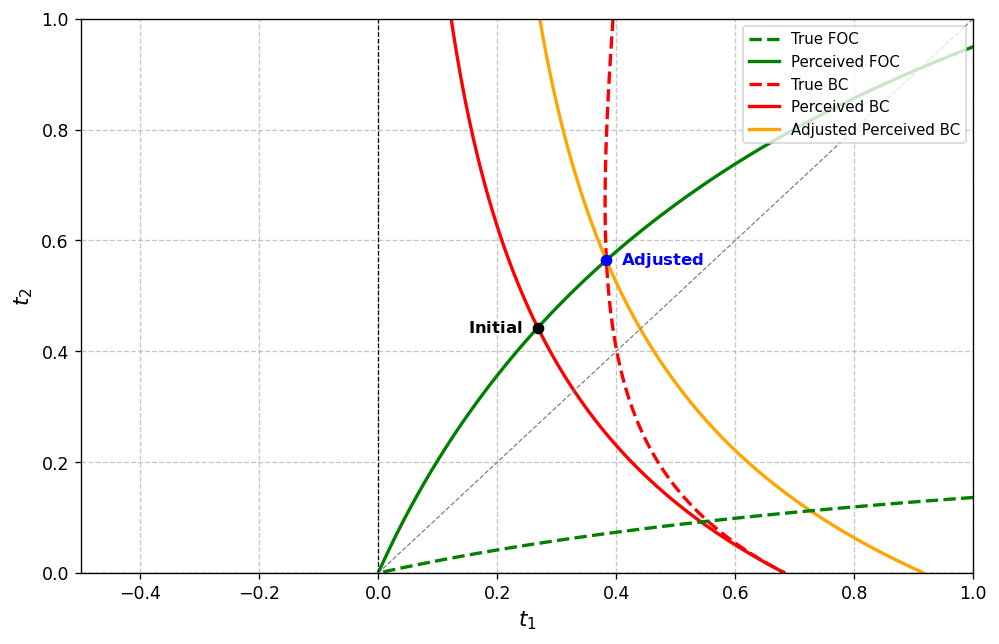}
\par\end{centering}
\caption{Case 1 ($e_1=0.60$, $e_2=2.50$, $\theta_2=0.55$, $R=0.50$)}\label{fig:fig1}
\end{figure}

Figure \ref{fig:fig1} visually summarizes the implications of overconfidence for commodity taxation in a two-good example. The red curves are government budget constraints. The solid red line is the set of tax rates satisfying equation (\ref{eq:percBC}): the \textit{perceived} budget constraint. The dashed red line represents the \textit{true} budget constraint, absent misperception.  The green curves depict first-order conditions for (perceived) welfare maximization. The solid green line is the set
of tax rates satisfying equation (\ref{eq:plannersFOC}):  the planner's \textit{perceived} FOC. The dashed green line represents the planner's \textit{true} FOC absent misperception.  

The pair of tax rates that solve the planner's perceived problem lie at the intersection of the solid red and solid green lines. This is marked as the \textit{Initial} solution. The example depicted in Figure \ref{fig:fig1} illustrates a straightforward case of the inverse-Ramsey result. Notice that, for the parameter values used here - $e_1=0.60$, $e_2=2.50$, and $\theta_2=0.55$ - not only is the elasticity on good 1 lower than that on good 2, but it is also lower than the perceived elasticity on good 2 ($\theta_2 e_2 =1.375  $). Thus, the ordering of perceived elasticities is the same as the ordering of true elasticities. Applying the standard Ramsey rule would therefore suggest that good 1 should be more highly taxed, and yet this is not what the overconfident decision-maker does. Rather, the \textit{Initial} solution imposes a higher tax rate on good 2, the relatively elastic good.

Finally, consider the solid orange curve in Figure \ref{fig:fig1}. This curve allows us to visualize the outcome of the budget adjustment rule discussed above. It depicts the perceived budget constraint for a new,higher revenue target: one that was selected so that the perceived planner's problem under this adjusted revenue target has a solution that falls on the true budget constraint. This is marked as the \textit{Adjusted} solution, and is found at the point where this orange curve intersects the perceived FOC (solid green curve). In this example, we observe that the inverse-Ramsey result persists even after the budget adjustment rule is implemented.

\begin{figure}
\begin{centering}
\includegraphics[width=0.8\paperwidth]{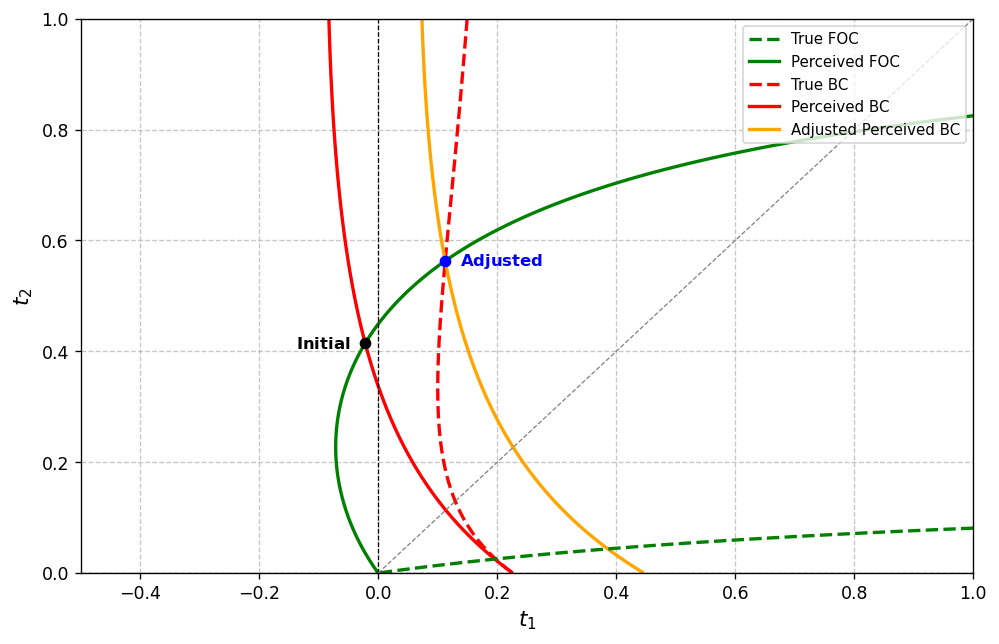}
\par\end{centering}
\caption{Case 2 ($e_1=0.60$, $e_2=4.00$, $\theta_2=0.45$, $R=0.20$)}\label{fig:fig2}
\end{figure}

While the result in Figure \ref{fig:fig1} is intriguing, it does not fully capture the degree to which this model can deviate from the standard Ramsey rule. In fact, it is even possible that the biased planner considers it preferable to \textit{subsidize} good 1. 
Figure \ref{fig:fig2} (parameterized with $e_1=0.60$, $e_2=4.00$, and $\theta_2=0.45$) depicts one such case.\footnote{In this specific instance, the solution after budget adjustment does not feature a subsidy, but this is not always the case.}

The subsidization case can help us build intuition. We can think of the model as describing the policy decisions of a representative voter who believes that their own demand for good 2 is relatively elastic as compared with some broader group of taxpayers. If this misperception becomes severe enough, eventually the voter will desire having a subsidy on good 1 financed by a large tax on good 2, because they expect that they will be able to avoid the tax at little cost, while other taxpayers will not. This same logic underpins all our inverse-Ramsey results, but it is particularly stark in the subsidy case.

In Appendix \ref{4cases}, we provide a more detailed characterization of the two types of inverse-Ramsey results discussed above, and present two additional cases.

\section{Other striking implications of overconfidence}

We have shown that a taxpayer/voter who is overconfident in a particular way about her ability to substitute away from taxed goods may prefer a commodity tax system that taxes elastically-demanded goods more heavily, in contradiction to the Ramsey rule for optimal commodity taxation. This occurs because the taxpayer perceives that the cost to her of taxing an elastic good relative to the budgetary cost is less than it really is. She also misperceives the extent to which a tax system achieves budget balance, but may still favor an inverse-Ramsey tax plan even if constrained to choose a budget-balancing set of commodity taxes.\footnote{It is worth noticing that one can also provide alternative interpretations
for the constrained maximization problem that we have set up in Section \ref{Section2}.
Consider, for instance, a heterogeneous-preferences setting where a proportion 
$\gamma$ of agents has preferences described by the
utility function $u\left( \mathbf{x},L\right) $, whereas the rest of the
population has preferences described by the function $v\left( 
\mathbf{x},L\right) $. Denote by $\mathbf{x}^{\ast }\left( \mathbf{q}
,w\right) $ and $\widetilde{\mathbf{x}}\left( \mathbf{q},w\right) $ the
demand vectors obtained maximizing, respectively, $u\left( \mathbf{x}
,L\right) $ and $v\left( \mathbf{x},L\right) $ subject to the private budget
constraint $\mathbf{q}\cdot \mathbf{x}=wL$; also, denote by $\overline{
\mathbf{x}}\left( \mathbf{q},w\right) $ the aggregate demand vector $\gamma 
\mathbf{x}^{\ast }\left( \mathbf{q},w\right) +\left( 1-\gamma \right) 
\widetilde{\mathbf{x}}\left( \mathbf{q},w\right) $. If $\gamma >1/2$, the
tax structure that arises as a political economy equilibrium is the one
favored by the agents whose preferences are described by $u\left( \mathbf{x}
,L\right) $, i.e. the outcome of the constrained maximization problem that
we have described in Section \ref{Section2}. Despite this isomorphism, one crucial
difference between the two settings is represented by the need for a
budget-adjustment rule. Under the model on which we have focused our
attention, i.e. the overconfident representative-agent model, a
budget-adjustment rule has to be ex-post implemented to guarantee that the
required revenue target is met. In the heterogeneous-preferences setting,
instead, no ex-post adjustment is needed.}

Overconfidence can lead to other strikingly unfamiliar tax prescriptions. We have seen that one such prescription is to favor a commodity tax system that features subsidies on some goods, a result that the Ramsey model cannot prescribe in the absence of cross-substitution elasticities. This occurs if large elasticity-overconfidence makes taxing a good so attractive -- raising revenue while inflicting little perceived own utility loss -- that a subsidy on another good is optimal.

Another remarkable implication of our results is that, in some cases, overconfidence can cause a distorting tax system to be preferred over a non-distorting lump-sum tax, even after the implementation of a budget adjustment rule. A formal analysis of this possibility is provided in Appendix \ref{lumpsum}. There, we show that this can only occur if, for some revenue target, the optimal tax structure features a subsidy on one good, as in Figure \ref{fig:fig2}. Intuitively, in this case, the overconfident agents believe they can shift so much of the burden of distortionary taxes onto others that they are better off under distortionary taxes than under lump-sum taxation.

In traditional optimal tax analysis with a representative agent, where minimizing excess burden is the only objective, lump-sum taxes are often ruled out\textbf{} because they are trivially superior, and ignoring them is often justified based on their unsuitability in a world of heterogeneous taxpayers and distributional considerations. Introducing overconfidence restores an optimal policy tradeoff between the perceived \textquotedblleft excess benefit\textquotedblright\ of commodity taxation and their distorting consequences.

\section{Conclusion}

Traditional optimal tax analysis, dating back almost a century to Ramsey (1927), prescribes that, other things equal, to minimize efficiency cost one should impose higher taxes on goods with lower demand elasticities. Yet policy makers rarely stress minimizing efficiency costs as a desideratum, and many commentators on policy seem to favor the opposite argument: that it is better to levy taxes on bases where taxpayer choice is higher.

In this note we revisit the commodity tax problem, and show that the attractiveness of the Ramsey inverse-elasticity prescription can itself be inverted if taxpayers who are overly confident of their ability to substitute away from taxed goods relative to others choose--or at least strongly influence--the tax system. A natural next step is to investigate what patterns of elasticity overconfidence consumers exhibit.

\appendix

\section{Appendix: Existence of Inverse-Ramsey Scenarios\label{existence}}

\renewcommand{\theequation}{A\arabic{equation}} \setcounter{equation}{0}

Suppose, without loss of generality, that ${e}_{i}>{e}_{j}$. In the standard model, the usual Ramsey inverse elasticity result implies that $ t_{j}> t_{i}$. We say that an inverse-Ramsey result holds for some perceived demand functions if the ordering of the perceived elasticities is the same as for the true elasticities ($\overline{e}_{i}>\overline{e}_{j}$), but the planner chooses tax rates that reverse the Ramsey rule ($ t_{j}< t_{i}$).

Consider a simple case where perceived demand for good $i$ under tax rate $ t_{i}>0$
is
\[
\bar{x}_{i}\left(t_{i}\right)\equiv\begin{cases}
x_{i}\left(t_{i}\right)^{\theta}x_{i}\left(0\right)^{1-\theta} & \text{if }t_{i}\geq0\\
x_{i}\left(t_{i}\right) & \text{if }t_{i}<0
\end{cases}.
\]
If good $i$ is subsidized, we assume demand is correctly perceived. But if the tax rate $t_i$ is positive, perceived demand is the geometric mean of realized demand for $i$, $x_{i}\left( t_{i}\right)$,
and (counterfactual) untaxed demand for $i$, $x_{i}\left(0\right)$. The parameter
$\theta$ controls the extent of overconfidence. When $\theta=1$,
demand for $i$ is correctly perceived. When $\theta=0$, the demand
for $i$ is (incorrectly) perceived as being perfectly inelastic. The perceived elasticity of demand for
good $i$ in this model is simply $\overline{e}_{i}=\theta{e}_{i}$. Thus, the smaller $\theta$ gets, the more the planner underestimates the responsiveness of demand for $i$ to taxation.

As in the main text, we assume that demand for good $j$ is accurately perceived.

It can be easily shown that there always exists some threshold value $\bar{\theta}$
such that $\overline{e}_{i}=\bar{\theta}{e}_{i}>{e}_{j}$ and an inverse-Ramsey result holds whenever $\theta<\bar{\theta}$. 

Recall, an inverse-Ramsey result holds whenever condition \eqref{condition} is satisfied. For any pair of tax rates we can always find some $\theta$
small enough that $\theta  =\frac{e_{j}\left( t_{j}\right)}{e_{i}\left( t_{i}\right)} $. In this case, condition \eqref{condition} simplifies to
\[
\frac{1-\frac{\alpha}{\mu}\left(\frac{x_{i}\left(t_{i}\right)}{x_{i}\left(0\right)}\right)^{1-\theta}}{1-\frac{\alpha}{\mu}}>1\iff x_{i}\left(0\right)>x_{i}\left( t_{i}\right),
\]
which is satisfied as long as true demand for good $i$ is not perfectly inelastic. By continuity of the planner's solution, condition \eqref{condition} must also hold at nearby values of $\theta$, including values where $\theta{e}_{i}>{e}_{j}$. Thus, there is always some value of $\theta$ small enough to deliver an inverse-Ramsey result. By the intermediate value theorem, there must exist some threshold value of $\overline{\theta }>e_{j}/e_{i}$ such that $\bar{\theta}{e}_{i}>{e}_{j}$, and an inverse Ramsey result is obtained for all $\theta<\bar{\theta}$.

We have thus demonstrated that an inverse-Ramsey result is always possible for some form of misperception, but providing a more precise characterization when it inverse-Ramsey result obtains is difficult due to the endogeneity of the various parameters included in condition \eqref{condition}. However, we can obtain an approximate characterization for small revenue requirements.

Consider a first-order Taylor approximation of equation \eqref{condition} with respect to the revenue requirement $R$, in the neighborhood of $R=0$. If---as we have assumed throughout the paper---good $i$ is correctly perceived when the tax rate is zero, this approximation reduces to\footnote{First, we index all endogenous parameters by the revenue requirement $R$ and define $\gamma\left(R\right)\equiv\frac{\mu\left(R\right)}{\alpha\left(R\right)}$. If demand is accurately perceived when tax rates are zero, then under a zero revenue requirement the planner's FOCs can be satisfied with a pair of zero tax rates, implying that $\gamma\left(0\right)=1$. Given this, note that the planner's FOC for good $i$ is approximated by
\begin{equation}
    -\gamma^{\prime}\left(0\right)-t_{i}^{\prime}\left(0\right)\left(\overline{e}_{i}\left(0\right)-e_{i}\left(0\right)\right)-t_{i}^{\prime}\left(0\right)\overline{e}_{i}\left(0\right)=0
\end{equation}
in the neighborhood of $R=0$. Importantly, this expression can be solved for $\gamma^{\prime}\left(0\right)$. Turning to equation \eqref{condition}, replacing left hand side with a local approximation, we obtain
\begin{equation}
e_{j}\left(0\right)\frac{\gamma^{\prime}\left(0\right)+\left(\overline{e}_{i}\left(0\right)-e_{i}\left(0\right)\right)t_i^{\prime}\left(0\right)}{\gamma^{\prime}\left(0\right)}>\overline{e}_{i}\left(0\right)
\end{equation}
substituting in the value of $\gamma^{\prime}\left(0\right)$ obtained earlier, we get the final approximation.}
\begin{equation*}
\frac{e_{i}+e_{j}}{2}>\overline{e}_{i}\label{eq:approxInvRam},
\end{equation*}
implying that, for small $R$, the inverse-Ramsey result holds whenever the perceived elasticity of demand for good $i$ is below the unweighted average of the (true) elasticities of good $i$ and good $j$.

\section{Appendix: Notes on Four Key Cases and Figure Analysis\label{4cases}}

\renewcommand{\theequation}{B\arabic{equation}} \setcounter{equation}{0}

\begin{figure}[H]
    \centering
    \includegraphics[width=0.8\paperwidth]{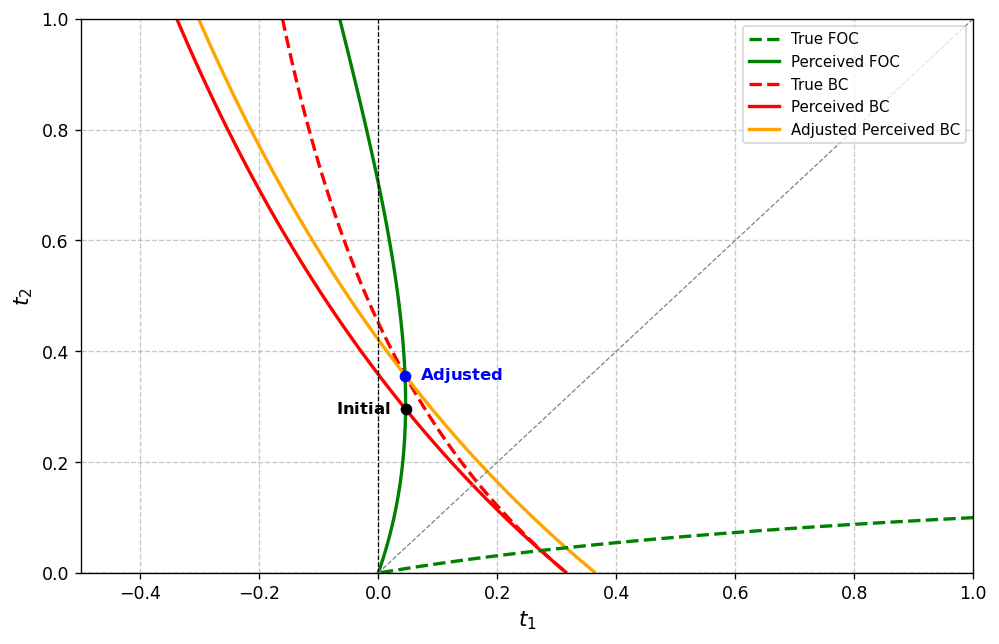}
    \caption{Case 3: Positive Slope, No Laffer Rate ($e_1=0.20$, $e_2=1.10$, $\theta_2=0.53$, $R=0.30$)}
    \label{fig:fig3}
\end{figure}

\begin{figure}[H]
    \centering
    \includegraphics[width=0.8\paperwidth]{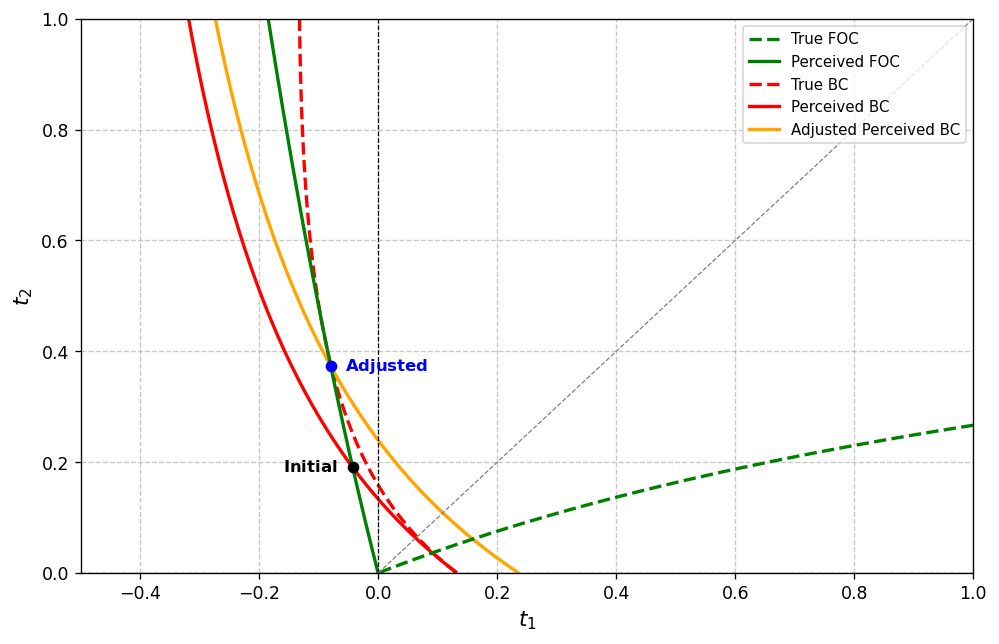}
    \caption{Case 4: Negative Slope, No Laffer Rate ($e_1=0.80$, $e_2=1.90$, $\theta_2=0.45$, $R=0.12$)}
    \label{fig:fig4}
\end{figure}

This appendix provides a more formal analysis of the cases illustrated in Figures \ref{fig:fig1} and \ref{fig:fig2}. In addition, we consider two more cases illustrating inverse-Ramsey results, depicted in Figures \ref{fig:fig3} and \ref{fig:fig4}. 

The four cases differ primarily in terms of the shape of the perceived first-order condition (FOC) curve. Below, we characterize the conditions under which each shape arises.

We consider the parametric model with two goods, $\theta_{1}=1$ (perfect perception for good 1), and $\theta_{2}\in\left[0,1\right]$ (misperception for good 2). The planner's perceived first-order condition (FOC), denoted as $F( t_{1}, t_{2})=0$, is given in log-form by:

\begin{equation}
F( t_{1}, t_{2})\equiv\log\!\biggl(1-e_{1}\frac{ t_{1}}{1+ t_{1}}\biggr)-(1-\theta_{2})e_{2}\log(1+ t_{2})-\log\!\biggl(1-\theta_{2}e_{2}\frac{ t_{2}}{1+ t_{2}}\biggr)=0
\end{equation}
This FOC defines the solid green curve in Figures \ref{fig:fig1}-\ref{fig:fig4} through $( t_{1}, t_{2})$-space.

\textbf{1. Initial Slope at the Origin}

To understand the behavior of the perceived FOC curve near the origin, we analyze its slope at $( t_{1}, t_{2})=(0,0)$.  Totally differentiating $F( t_{1}, t_{2})=0$ and evaluating at the origin yields the initial slope:

\begin{equation}
\left.\frac{\mathrm{d} t_{1}}{\mathrm{d} t_{2}}\right|_{0,0}=2\cdot\frac{e_{2}}{e_{1}}\cdot\left(\theta_{2}-\frac{1}{2}\right).
\end{equation}

The sign of this initial slope is crucial in determining the overall shape of the perceived FOC curve:

\begin{itemize}
    \item \textbf{Positive Initial Slope:}  Occurs when $\theta_{2}>\frac{1}{2}$. In this case, at very low tax rates, the perceived optimal policy involves increasing both $ t_1$ and $ t_2$ in tandem.
    \item \textbf{Negative Initial Slope:}  Occurs when $\theta_{2}<\frac{1}{2}$. Here, starting from zero taxes, the perceived optimal policy initially involves a trade-off: increasing $ t_2$ requires decreasing $ t_1$ to maintain the perceived optimality condition.
\end{itemize}

\textbf{2. Existence of a Positive Laffer Peak (Vertical Slope)}

The perceived FOC curve can exhibit a backwards-bending shape, indicated by a vertical slope. This occurs when $\frac{\partial F}{\partial t_{2}}=0$, which simplifies to:

\begin{equation}
\frac{\theta_{2}}{1+\left(1-\theta_{2}e_{2}\right) t_{2}}=1-\theta_{2}.
\end{equation}

Solving for $ t_{2}$ gives the tax rate at which the vertical slope potentially occurs:

\begin{equation}
 t_{2}^{vertical}=\frac{2\left(\theta_{2}-\frac{1}{2}\right)}{(1-\theta_{2})(1-\theta_{2}e_{2})}.
\end{equation}

For a vertical slope to occur at a \textit{positive} tax rate ($ t_{2}^{vertical}>0$), we require the numerator and denominator to have the same sign. Since $(1-\theta_{2})$ is always positive for $\theta_2 \in [0, 1)$, the sign is determined by the numerator and $(1-\theta_{2}e_{2})$.

\begin{itemize}
    \item \textbf{Condition for Positive $ t_{2}^{vertical}$:} $\frac{2\left(\theta_{2}-\frac{1}{2}\right)}{(1-\theta_{2})(1-\theta_{2}e_{2})}>0$.  Notice the numerator shares the same sign as the initial slope condition. The denominator, $(1-\theta_{2}e_{2})$, is positive if and only if there is no perceived Laffer rate for good 2 ($\theta_{2}e_{2}<1$). However, for a positive $ t_{2}^{vertical}$ and given that $(1-\theta_2) > 0$, we need $\theta_2 > 1/2$ and $(1-\theta_2 e_2) > 0$, or $\theta_2 < 1/2$ and $(1-\theta_2 e_2) < 0$.
\end{itemize}

\textbf{3. Shapes of the Solution Curve and Case Breakdown (Figures \ref{fig:fig1}-\ref{fig:fig4})}

The interplay between the initial slope and the existence of a positive Laffer peak, determined by the parameter $\theta_2$ relative to $1/2$ and $1/e_2$, leads to four distinct cases, illustrated in Figures \ref{fig:fig1}-\ref{fig:fig4}:

\begin{itemize}
    \item \textbf{Case 1: Figure \ref{fig:fig1} :}  \textit{Positive slope at the origin, and there exists a perceived Laffer rate on good 2}. 
    \[
    \theta_{2}>\frac{1}{2},\quad\text{and}\quad\theta_{2}e_{2}>1.
    \]
    Notice, this implies that
    \[
    \theta_{2}>\max\left\{ \frac{1}{2},\frac{1}{e_{2}}\right\} .
    \]
    \item \textbf{Case 2: Figure \ref{fig:fig2}:} \textit{Negative slope at the origin, and there exists a perceived Laffer rate on good 2}. 
    \[
    \theta_{2}<\frac{1}{2},\quad\text{and}\quad \theta_{2}e_{2}>1.
    \]
    Notice, this implies that
    \[
    \theta_{2}\in\left(\frac{1}{e_{2}},\frac{1}{2}\right)
    \]
    and thus, this case can only occur if $e_{2}>2$.
    \item \textbf{Case 3: Figure \ref{fig:fig3}:} \textit{Positive slope at the origin, and no Laffer rate on good 2}.
    \[
    \theta_{2}>\frac{1}{2},\quad\text{and}\quad 1>\theta_{2}e_{2}.
    \]
    Notice, this implies that
    \[
    \theta_{2}\in\left(\frac{1}{2},\frac{1}{e_{2}}\right),
    \]
    and thus, this case can only occur if $e_{2}<2$.
    \item \textbf{Case 4: Figure \ref{fig:fig4} :} \textit{Negative slope at the origin, and no Laffer rate on good 2}.
    \[
    \theta_{2}<\frac{1}{2},\quad\text{and}\quad1>\theta_{2}e_{2}.
    \]
    Notice this implies that
    \[
    \theta_{2}<\min\left\{ \frac{1}{2},\frac{1}{e_{2}}\right\} .
    \]
\end{itemize}
This analysis demonstrates how the degree of overconfidence, parameterized by $\theta_2$, and the relative elasticities of demand, $e_1$ and $e_2$, interact to determine the shape of the perceived optimal policy space and the resulting tax policy outcomes, ranging from strong inverse-Ramsey results to outcomes closer to the traditional Ramsey rule.

\section{Appendix: Commodity Taxation versus Lump-Sum Taxation\label{lumpsum}}

\renewcommand{\theequation}{C\arabic{equation}} \setcounter{equation}{0}

Consider the set of first-order conditions that characterize an optimal
commodity tax structure in a standard representative agent setting without
overconfidence. For $i$ = 1,2 (and assuming for simplicity a quasi-linear framework, as in our example, so that the private marginal utility of
income, $\alpha $, is constant and equal to 1), these are given by
\begin{equation}
-1+\mu \left( 1-\frac{t_{i}}{1+t_{i}}e_{i}\right) =0.  \label{j}
\end{equation}

Due to the excess burden of commodity taxation (reflected in the elasticity
term), both $t_{1}$ and $t_{2}$ would necessarily be positive, and the
Lagrange multiplier $\mu $ would be greater than $1$. With lump-sum
(hereafter, LS) taxation, instead, there would be no excess burden and $\mu $
would be equal to $1$. Taking into account that $\mu $ represents the
utility loss endured by a representative agent when the revenue target is
marginally increased, LS taxation is preferred over commodity taxation
because each agent realizes that the private cost of raising one extra
dollar for the public budget through a commodity tax on good $i$ is larger
than the private cost sustained under a uniform LS tax that raises one extra
dollar: formally, $1-\frac{t_{i}}{1+t_{i}}e_{i}<1$ whenever $t_{i}>0$.

In our setting with overconfidence, and under the assumption that $\theta
_{1}=1$ and $\theta _{2}<1$, the set of first-order conditions becomes
\begin{eqnarray}
-1+\mu \left( 1-\frac{t_{1}}{1+t_{1}}e_{1}\right) &=&0,  \label{pizza} \\
-1+\mu \frac{\overline{x}_{2}}{x_{2}}\left( 1-\frac{t_{2}}{1+t_{2}}\theta
_{2}e_{2}\right) &=&0.  \label{pasta}
\end{eqnarray}

The fact that $\theta _{2}<1$ implies that $\frac{\overline{x}_{2}}{x_{2}}>1$
when $t_{2}>0$; thus, one cannot rule out the possibility that, although $1-
\frac{t_{2}}{1+t_{2}}\theta _{2}e_{2}<1$ when $t_{2}>0$, the value of the
product $\frac{\overline{x}_{2}}{x_{2}}\left( 1-\frac{t_{2}}{1+t_{2}}\theta
_{2}e_{2}\right) $ in (\ref{pasta}) is larger than one. The reason is that,
even though the agent takes into account (at least partially, since $\theta
_{2}<1$) that an increase in $t_{2}$ comes with a behavioral
response that dampens its positive revenue effects ($1-\frac{t_{2}}{1+t_{2}}
\theta _{2}e_{2}<1$), she overestimates the revenue gains due to the
mechanical effect of raising $t_{2}$. In fact, whereas all agents consume an
amount $x_{2}\left( t_{2}\right) $ of good 2 (prior to the increase in $t_{2}
$), each believes that all the others, having a less elastic demand, are
consuming an amount $\overline{x}_{2}\left( t_{2}\right) >x_{2}\left(
t_{2}\right) $ (for any positive value of $t_{2}$). Thus, if the ratio $
\frac{\overline{x}_{2}}{x_{2}}$ is sufficiently large (so that $\frac{
\overline{x}_{2}}{x_{2}}\left( 1-\frac{t_{2}}{1+t_{2}}\theta
_{2}e_{2}\right) >1$), an overconfident (henceforth OC) agent believes that the private cost of raising
one extra dollar for the public budget, via an increase in $t_{2}$, is lower
than the private cost that she would sustain under a uniform lump-sum tax
that raises one extra dollar. Intuitively, whereas an OC agent understands
that a uniform LS tax hits everybody symmetrically, she fails to realize
that the same will be true when good 2 is taxed;\ in the latter case, she
believes that the tax burden will fall disproportionately on the other
taxpayers.

To assess whether commodity taxation is regarded as better than LS taxation
for a given revenue target $R$, what matters is the integral $
\int_{0}^{R}\mu \left( r\right) dr$, where $\mu \left( r\right) $ represents
the value of $\mu $ when the revenue requirement is $r$. This integral represents the total utility loss anticipated by a representative agent
when the revenue extracted from the taxpayers is gradually raised all the way
to the target $R$. Commodity taxation
is perceived as better than LS taxation when $\int_{0}^{R}\mu \left(
r\right) dr$ is lower under commodity taxation than under LS taxation. In
our example (where $\alpha =1$ due to quasi-linearity), this condition
requires that $\int_{0}^{R}\mu \left( r\right) dr$, evaluated under
commodity taxation, is smaller than $R$ (the utility loss
suffered by an agent under a lump-sum tax $T=R$). Denote by $\overline{\mu }
^{CT}\left( R\right) $ the value of $\int_{0}^{R}\mu \left( r\right) dr$
when revenue is raised through commodity taxation. In order to
have $\overline{\mu }^{CT}\left( R\right) <R$, a necessary condition\ is
that, within the range $r\in (0,R]$, there are at least some values of $r$
for which $\mu \left( r\right) <1$. From the first-order
conditions (\ref{pizza})-(\ref{pasta}), one can easily see that, when they are jointly satisfied for $\mu <1$, it must necessarily be that $t_{1}<0$ and 
$t_{2}>0$.\footnote{Intuitively, when $\mu <1$, the burden privately experienced from taxing
good $2$ is felt to be so light, relatively to the revenue that it allows
collecting from each of the other taxpayers, that an OC agent has the
incentive to set $t_{2}$ at a rate that, based on her beliefs, generates
more revenue than the target $R$. In this way, the extra-revenue can be
rebated back via a subsidy on good $1$. Even though an OC agent does not
expect her consumption of good $1$ to be larger than that of her
fellow taxpayers (and therefore realizes that her gain from the subsidy
will be no different than other taxpayers' gain), she favors such a tax
arrangement due to the belief that the financing of the subsidy falls
disproportionately on other people.} Thus, in order to have $\overline{\mu }^{CT}\left( R\right) <R$,
a necessary condition is that, within the set $(0,R]$, 
there are at least some revenue targets for which an OC agent would choose to tax good 2
and subsidize good 1.

Finally, notice that, under the assumptions made in our illustrative example
of Section 3, the ratio $\frac{\overline{x}_{2}}{x_{2}}$ is given by$\left(
1+t_{2}\right) ^{\left( 1-\theta _{2}\right) e_{2}}$, which is monotonically
increasing in $t_{2}$ and takes the value $1$ at $t_{2}=0$. Given that, in order
for (\ref{pasta}) to be satisfied with $\mu <1$, it must be that $\frac{
\overline{x}_{2}}{x_{2}}\left( 1-\frac{t_{2}}{1+t_{2}}\theta
_{2}e_{2}\right) >1$, it follows that commodity taxation cannot be regarded
as better than LS taxation when%
\begin{equation}
\partial \left[ \left( 1+t_{2}\right) ^{\left( 1-\theta _{2}\right)
e_{2}}\left( 1-\frac{t_{2}}{1+t_{2}}\theta _{2}e_{2}\right) \right]
/\partial t_{2}<0\text{ \ \ for }t_{2}\geq 0.  \label{aperitivo}
\end{equation}

It is easy to verify that (\ref{aperitivo}) holds when $\theta
_{2}\geq 1/2$ and $\theta _{2}e_{2}>1$, i.e. provided that the degree of
elasticity-misperception is sufficiently small and an OC agent perceives
the existence of a Laffer rate for $t_{2}$. However, when either one
or both of these conditions are not satisfied, one can always provide
examples where an OC agent prefers commodity taxation over LS taxation.
\end{document}